\documentclass[twocolumn,aps,prl]{revtex4-1}

	\usepackage[usenames,dvipsnames]{xcolor}
	\usepackage{amsmath}
	\usepackage{amsfonts}
	\usepackage{amssymb}
	\usepackage{bm}
	\usepackage[colorlinks=true,citecolor=blue,linkcolor=red]{hyperref}
	\usepackage{graphicx}
	\usepackage{bbold}					
	\usepackage[makeroom]{cancel}		
	\usepackage{multirow}				
	\usepackage[normalem]{ulem}        
	\usepackage{array}
	\usepackage{pdfpages}
	\makeatletter
	\AtBeginDocument{\let\LS@rot\@undefined}
	\makeatother
	
	\newcolumntype{x}[1]{>{\centering\let\newline\\\arraybackslash\hspace{0pt}}p{#1}}

	\DeclareMathOperator{\sign}{sign}  	

	\DeclareMathAlphabet{\mathbbold}{U}{bbold}{m}{n}

	\newcounter{subeqn} %
	\makeatletter
	\@addtoreset{subeqn}{equation}
	\makeatother

\definecolor{TB}{rgb}{0,0,0} 

\begin{document}

\title{Jones polynomial and knot transitions in topological semimetals}

\author{Zhesen Yang$^{1,2}$}
\author{Ching-Kai Chiu$^{3}$}\email{qiujingkai@ucas.edu.cn}
\author{Chen Fang$^{1}$}\email{cfang@iphy.ac.cn}
\author{Jiangping Hu$^{1,3,4}$}\email{jphu@iphy.ac.cn}

\affiliation{$^{1}$Beijing National Laboratory for Condensed Matter Physics,
	and Institute of Physics, Chinese Academy of Sciences, Beijing 100190, China}
\affiliation{$^{2}$University of Chinese Academy of Sciences, Beijing 100049, China}
\affiliation{$^{3}$Kavli Institute for Theoretical Sciences and CAS Center for Excellence in Topological Quantum Computation, University of Chinese Academy of Sciences, Beijing 100190, China}
\affiliation{$^{4}$South Bay Interdisciplinary Science Center, Dongguan, Guangdong Province, China}

\date{\today}

\begin{abstract}
Topological nodal line semimetals host stable chained, linked, or knotted line degeneracies in momentum space protected by symmetries. In this paper, we use the Jones polynomial as a general topological invariant to capture the global knot topology of the nodal lines. We show that every possible change in Jones polynomial is attributed to the local evolutions around every point where two nodal lines touch. As an application of our theory, we show that nodal chain semimetals with four touching points can evolve to a Hopf-link. We extend our theory to 3D non-Hermitian multi-band exceptional line semimetals.
\end{abstract}


\maketitle

{\em Introduction---}Topological phases of matter have been attracting extensive attention in the field of condensed matter physics \cite{HassanKane2010,QiZhang2011,Review_TI_Das,RevModPhys.88.035005,review_semimetal_a,reveiw_semimetal_b}. Although the topological invariants of gapped phases are defined globally, they can be locally analyzed by studying the low energy theories of some gapless points in the Brillouin zone (BZ) from a critical phase~\cite{BernevigHughesBook13,Taylor}. For example, the Chern number can be calculated by analyzing the mass terms around all the Dirac points~\cite{BernevigHughesBook13,Bernevig:2006kx}. In this sense, all the gapped phases can be generated from those critical gapless phases by adding different types of perturbations~\cite{Nontrivial_surface_chiu,RevModPhys.88.035005}. 

The topological nodal line semimetals protected by chiral symmetry or space-time inversion symmetry can host stable one-dimensional (1D) degeneracy line in the 3D BZ \cite{reveiw_semimetal_b}. These nodal lines can form loops \cite{HoravaPRL05, BurkovBalentsPRB11,Chen:2015aa, Kane_Cu3N_ring,PhysRevLett.115.026403,Dai_Cu3PdN_ring,Nodal_Line_Fang,Bian:2016aa,PhysRevB.93.205132,Schoop:2016aa},  chains~\cite{Bzdusek:2016aa,PhysRevLett.119.036401,Yan:2018aa,PhysRevMaterials.2.014202,PhysRevLett.120.106403, PhysRevX.8.031044, PhysRevB.98.161104, Lou:2018aa, PhysRevB.97.201107, PhysRevB.98.075146, PhysRevB.98.085122, 2018arXiv180807469W,PhysRevB.99.045130,PhysRevB.99.075131,Lian:2019aa}, links \cite{PhysRevB.96.041102,PhysRevB.96.041103, PhysRevB.96.081114, PhysRevB.97.155140, PhysRevB.98.205410,PhysRevLett.119.156401,PhysRevLett.119.147001, PhysRevLett.121.036401,PhysRevB.96.041202} or  knots \cite{PhysRevB.96.041202, PhysRevB.96.201305}. Their topological properties are not only captured by the local charge \cite{reveiw_semimetal_b} but also described by the global knot invariant \cite{PhysRevLett.119.147001,PhysRevB.95.094512,PhysRevB.96.155105,2019arXiv190300018T,PhysRevB.99.081102}. Two nodal knot semimetals (hereafter, {\em knot} refers both {\em link} and {\em knot}) belong to the same (topological equivalence) classes, if their nodal lines can be deformed to each other by non-broken bending and stretching without crossing each other~\cite{KnotsTheoryandPhysics}. Being analogue to a Dirac point as a topological phase transition, a touching point (TP), where two nodal lines touch together, might be a knot transition between two distinct knot classes. 
If we start from this critical phase, by adding different types of symmetry allowed perturbations, as the TPs are removed, different trivial and  nontrivial nodal knot semimetals can be generated, which are dubbed as {\em generated phases}.
A question naturally arises whether the knot topology of the nodal lines can be characterized by analyzing the local evolutions around TPs. 
The answer to this question provides a guide to analyze the possible generated phases emerging from nodal chain semimetals~\cite{Bzdusek:2016aa,PhysRevLett.119.036401,Yan:2018aa,PhysRevMaterials.2.014202,PhysRevLett.120.106403, PhysRevX.8.031044, PhysRevB.98.161104, Lou:2018aa, PhysRevB.97.201107, PhysRevB.98.075146, PhysRevB.98.085122, 2018arXiv180807469W,PhysRevB.99.045130,PhysRevB.99.075131,Lian:2019aa}, which are symmetry protected critical phases with multiple TPs. 

\begin{figure}[t]
\centerline{\includegraphics[height=3.4cm]{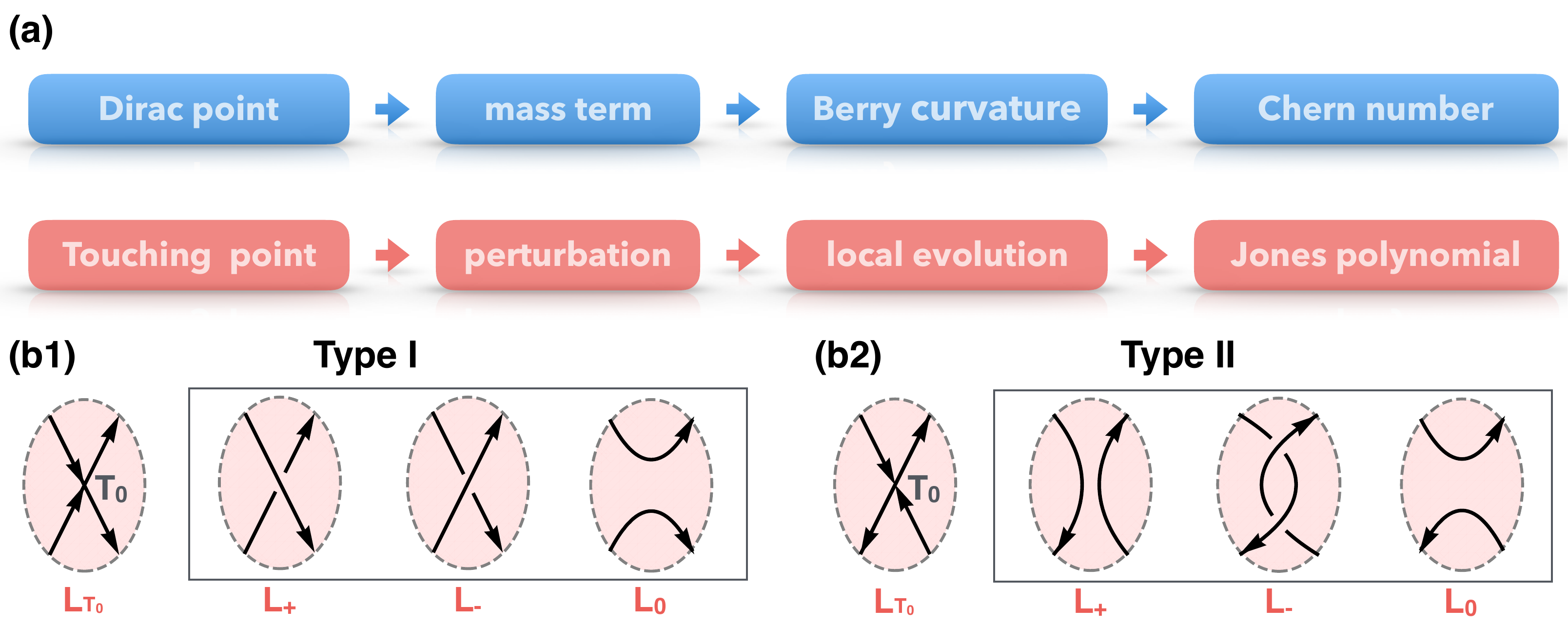}}
\caption{The global topological invariants can be characterized by the local physics around some special points. The Chern insulators (nodal knot semimetals) can be viewed as generated phases from a critical phase with one or several Dirac points (TPs). The transition of Chern number (Jones polynomial) is attributed to the evolution changes of Berry curvature (local nodal lines) around the Dirac points (TPs) in the presence of perturbations.   (b) shows two possible line orientations (type I/II) around the TPs and the arrows indicate the directions of the nodal lines. There are three possible local evolutions for each line orientation. 
\label{F1}}
\end{figure}

\begin{figure*}[t]
\centerline{\includegraphics[height=2.86cm]{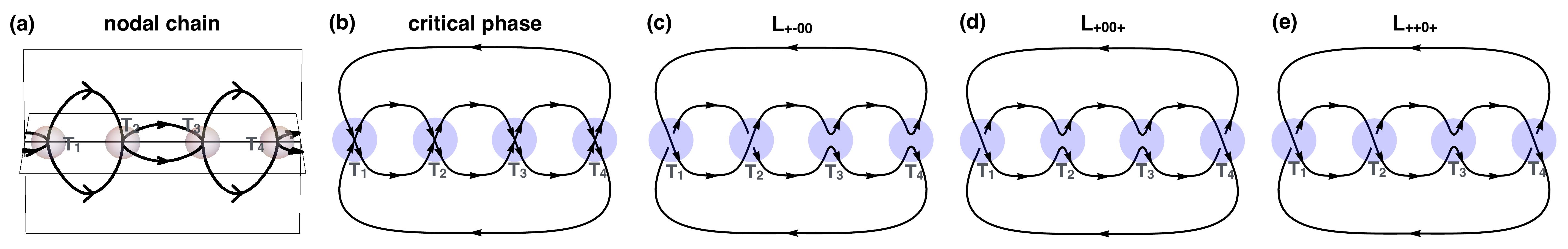}}
\caption{Critical phase with TPs and generated phases without TPs. (a) shows the nodal chain semimetal with 4 TPs. (b) shows the critical phase, which is equivalent to (a) based on the periodic boundary condition of the 3D BZ. (c)-(e) show several examples of the generated phases after the local evolutions of the 4 TPs.  
\label{F2} }
\end{figure*}

In this paper, we first show that  the Jones polynomial \cite{KnotsTheoryandPhysics,Knotnote,SM1} can faithfully characterize the orientated nodal knot semimetals protected by chiral symmetry. Similar to the transition of Chern insulator, we also show that the transition of Jones polynomial can be analyzed by studying the local evolutions around all the TPs in a critical phase as shown in Fig.~\ref{F1}. Furthermore, the low energy theory of the TP provides an additional constraint to the local evolutions and excludes the emergence of some generated phases. To demonstrate this theory, we use a nodal chain semimetal with four TPs as an example to show all of the possible generated phases, such as the emergence of a Hopf-link with the non-zero linking number~\cite{PhysRevB.96.041103}. In the end, we extend the recipe of the knot topology analysis to non-Hermitian exceptional line semimetals~\cite{PhysRevLett.118.045701,PhysRevB.97.075128,PhysRevA.98.042114,PhysRevB.99.081102,PhysRevB.99.161115,PhysRevB.99.041116,PhysRevB.99.161115,wangNonHermitianNodallineSemimetals2019b,PhysRevLett.123.066405}, since Hermitian chiral symmetric systems and non-Hermitian systems share the identical mathematical structures. 

{\em Nodal line semimetals protected by chiral symmetry---}We start with a general $2N-$bands Bloch Hamiltonian preserving chiral symmetry 
\begin{equation}\begin{aligned}
\mathcal{H}_0(\bm{k})=h_0(\bm{k})\tau_++h_0^\dag(\bm{k})\tau_-, 
\end{aligned}
\label{H0}\end{equation}
where $\tau_\pm=(\tau_x\pm i\tau_y)/2$, $h_0(\bm{k})$ is an $N\times N$ matrix, and chiral symmetry operator $\mathcal{S}=\tau_z$. Due to chiral symmetry, the Hamiltonian obeys $S\mathcal{H}_0(\bm{k})S^{-1}=-\mathcal{H}_0(\bm{k})$. Since $\det[\mathcal{H}_0(\bm{k})]$ is the product of all the energies, the locations of the nodal lines at $E=0$ are determined by  
\begin{equation}
\det[h_0(\bm{k})]=\det[h_0^\dag(\bm{k})]^*=\gamma_0^r(\bm{k})+i\gamma_0^i(\bm{k})=0,
\label{E1}
\end{equation}
and chiral symmetry leads to at least 2-fold degeneracy in the nodal lines. 
In other words, the two constraints of $\gamma_0^r(\bm{k})=0$ and $\gamma_0^i(\bm{k})=0$ determine two surfaces in the 3D BZ respectively so that their crossings form the nodal lines. 

A topological invariant characterizing each individual nodal line is given by the winding number~\cite{SM1}
\begin{equation}
\nu=\frac{i}{2\pi}\oint_{\Gamma(\bm{K_0})}d\bm{k} \cdot \nabla_{\bm{k}} \left(\ln\det[h_0(\bm{k})]\right),
\label{WN}
\end{equation}
where $\bm{K}_0$ is a point located at the nodal lines and $\Gamma(\bm{K}_0)$ is a closed-loop enclosing the nodal line and centered at $\bm{K}_0$. If the winding number is non-zero, the integral path is not contractible so that these nodal lines are topologically protected and can not be gapped in the presence of any weak chiral-symmetric perturbations. Since the winding number can change its sign under the reverse of the integral path, the direction of the nodal line is given by the normal vector of the oriented integral path corresponding to the positive winding number. By marking an arrow pointing this normal direction along the nodal line aligned with the counterclockwise integral path, one can assign an orientation to these nodal lines. 

{\em Critical phase and generated phases---} Now we show that all the generated phases from a critical phase with perturbations can be determined by local evolutions around every TP. Consider that the line node system is in a critical phase with $m$-TPs, which can be labeled by $L_{T_1...T_m}$, as shown in Fig. \ref{F2} (a) with $m=4$. Then we can ask the following question: by adding a general form of perturbation respecting chiral symmetry 
\begin{equation}
\mathcal{H}_1(\bm{k},\lambda)=\lambda h_{1}(\bm{k},\lambda)\tau_++\lambda h_{1}^\dag(\bm{k},\lambda)\tau_-,
\label{E2}
\end{equation}
where 
$\lambda$ is an external parameter, what type of a nodal knot as a generated phase can be generated by the perturbation?
Since the perturbation is weak, only the local evolutions around the TPs finally determine the linking or knotting properties of the nodal line.
To systematically study the generated phases, we project the critical phase in the 3D BZ (Fig.~\ref{F2}(a)) into a 2D plane and deform the projection to the diagram in Fig.~\ref{F2}(b) based on the periodic boundary condition of the 3D BZ. According to the directions of the nodal lines near the TPs, there exist two different types of TPs and the corresponding local evolutions $L_{0/+/-}$, namely type I/II TP and type I/II local evolutions as shown in Fig.~\ref{F1}(b1)/(b2)~\cite{SM1}. In this regard, the generated phases evolving from the critical phase (multiple TPs) can be labeled by $L_{n_1...n_m}$, where $n_i=0,\pm$ represent the local evolutions near the i-th TP $T_i$;  Fig.~\ref{F2}(c)-(e) show several possible generated phases. We note that $L_{n_1...n_m}$ are the 2D projection representation of the 3D knots, which is known as knot diagram~\cite{SM1}. Although different projection planes lead to distinct knot diagrams of the same knot, the invariant, which will be given later, is independent of the choice of the projection plane~\cite{KnotsTheoryandPhysics}. 

{\em Jones polynomial---} Having obtained all the perturbation generated phases, we define the corresponding knot invariant to classify them. In knot theory, the topology of inequivalent knots can be distinguished by distinct knot  polynomials~\cite{KnotsTheoryandPhysics}. We specifically use the Jone polynomial $J(L_{\#})$ to characterize knots $L_{\#}$ in the nodal line  semimetals. The reason to choose this polynomial is that the Jones polynomial can distinguish the orientations of the knots~\cite{Jones:1985aa} from the directions of the winding numbers as well as reveals that the knot topology connects potential physical observables by using Chern-Simons theory~\cite{Witten1989,PhysRevB.95.094512,Frohlich:1989aa}. Mathematically, the Jones polynomials~\cite{KnotsTheoryandPhysics} is a Laurent polynomial $\mathbb{Z}[t^{1/2},t^{-1/2}]$, which satisfies (i) the so-called skein relation
\begin{equation}
t^{-1} J\left(L_{+}\right)-t J\left(L_{-}\right)+\left(t^{-1 / 2}-t^{1 / 2}\right) J\left(L_{0}\right)=0,
\label{JP}
\end{equation}
where $L_+$, $L_-$ and $L_0$ are three oriented knots that are identical except in the small red region as shown in Fig.~\ref{F1}(b1); (ii) initial condition 
$J(O)\equiv 1$, where $O$ represents an unknot (ring). Any two equivalent orientated knots have the same Jones polynomial. Based on the definition, the skein relation only relates the Jones polynomials of type I local evolution around type I TP. After extending the skein relation from type I to type II (Fig. \ref{F2} (b2))~\cite{SM1}, we can connect all the generated phases by using Eq.~\ref{JP}. Hence, $J(L_{n_1...n_m})$ can be calculated systematically via the skein relation and the initial condition $J(O)=1$.

\renewcommand\arraystretch{1.5}
\begin{table}
\caption{\label{character2} The Jones polynomials represent all the generated phases $L_{n_1n_2n_3n_4}$ evolving from the nodal chain in Fig. \ref{F2} (a), where $n_i=0,\pm1$. As $\vec{\nabla}_{\bm{k}} \det [ h_0(\bm{k})]\neq 0$ at each TP, the evolution is limited to the three possible generated phases marked by the red color. 
}
\label{T1}
\begin{tabular*}{8.5cm}{|p{2cm}|p{4cm}|p{2cm}|}
  \hline
Links or knots & Jones polynomial & $n=\sum_{i=1}^4n_i$  \\\hline 
\noindent{\color{red}Unknot} & \noindent{\color{red}1} & \noindent{\color{red}$|n|=1$}\\\hline
\noindent{\color{red}Unlink} & \noindent{\color{red}$-t^{-1/2}-t^{1/2}$}  & \noindent{\color{red}$|n|=0$}  \\\hline
\noindent{\color{red}Hopf-link} & \noindent{\color{red} $-t^{5\sign(n) /2}-t^{\sign(n)/2}$} & \noindent{\color{red}$|n|=2$} \\\hline
Trefoil knot & $-t^{4\sign(n)}+t^{3\sign(n)}+t^{\sign(n)}$ & $|n|=3$ \\\hline
Solomon's knot & $-t^{9\sign(n)/2}-t^{5\sign(n)/2}+t^{3\sign(n)/2}-t^{\sign(n)/2}$ & $|n|=4$  \\\hline
\end{tabular*}
\end{table}

To demonstrate the approach of obtaining the explicit form of the Jones polynomial, we consider the evolution of the nodal chains with 4 TPs in the semimetals as shown in Fig.~\ref{F2} (a). 
This nodal chain semimetal is used as an example through this manuscript. Due to the orientations of the nodal lines, each local evolution near the TP can transit to three configurations $L_+,L_-$ and $L_0$ of type I in Fig.~\ref{F1}(b1). First, we start with two unknots $L_{+-0+}$ and $L_{+-0-}$ with $J(L_{+-0+})=J(L_{+-0-})=1$. The skein relation (\ref{JP}) at TP $T_4$ connects the two unknots and the unlink (two separated loops) $L_{+-00}$ as shown in Fig.~\ref{F2}(c); therefore, $J(L_{+-00})=-t^{-1/2}-t^{1/2}$. Secondly, knowing the Jones polynomials of the unlink and the unknot, we have $J(L_{+00-})=-t^{-1/2}-t^{1/2}$ and $J(L_{+000})=1$ and then obtain the polynomial of the Hopf-link $J(L_{+00+})=-t^{5/2}-t^{1/2}$ by the skein relation at $T_4$ as illustrated in Fig.~\ref{F2}(d). 
Thirdly, the skein relation at $T_2$ also connects an unknot $L_{+-0+}$, a Hopf-link $L_{+00+}$, and a trefoil knot $L_{++0+}$ as shown in Fig.~\ref{F2}(e); hence, trefoil knot invariant is given by $J(L_{++0+})=-t^4+t^3+t$. By following these rules, we can have the Jone polynomials for the $3^4$ configurations of $L_{n_1n_2n_3n_4}$ listed in Table \ref{T1}. In this model, the topology of the generated phases 
can be simply determined by the summation of the local diagram around every TP, which is similar to the transition of Chern number as shown in Fig.~\ref{F1} (a). By using this example, it is not difficult to extend our analysis to any generic critical phase with several TPs. Only the local evolution around the TP plays an essential role in determining the topology of generated phases. 

\begin{figure}[t]
\centerline{\includegraphics[height=3.9cm]{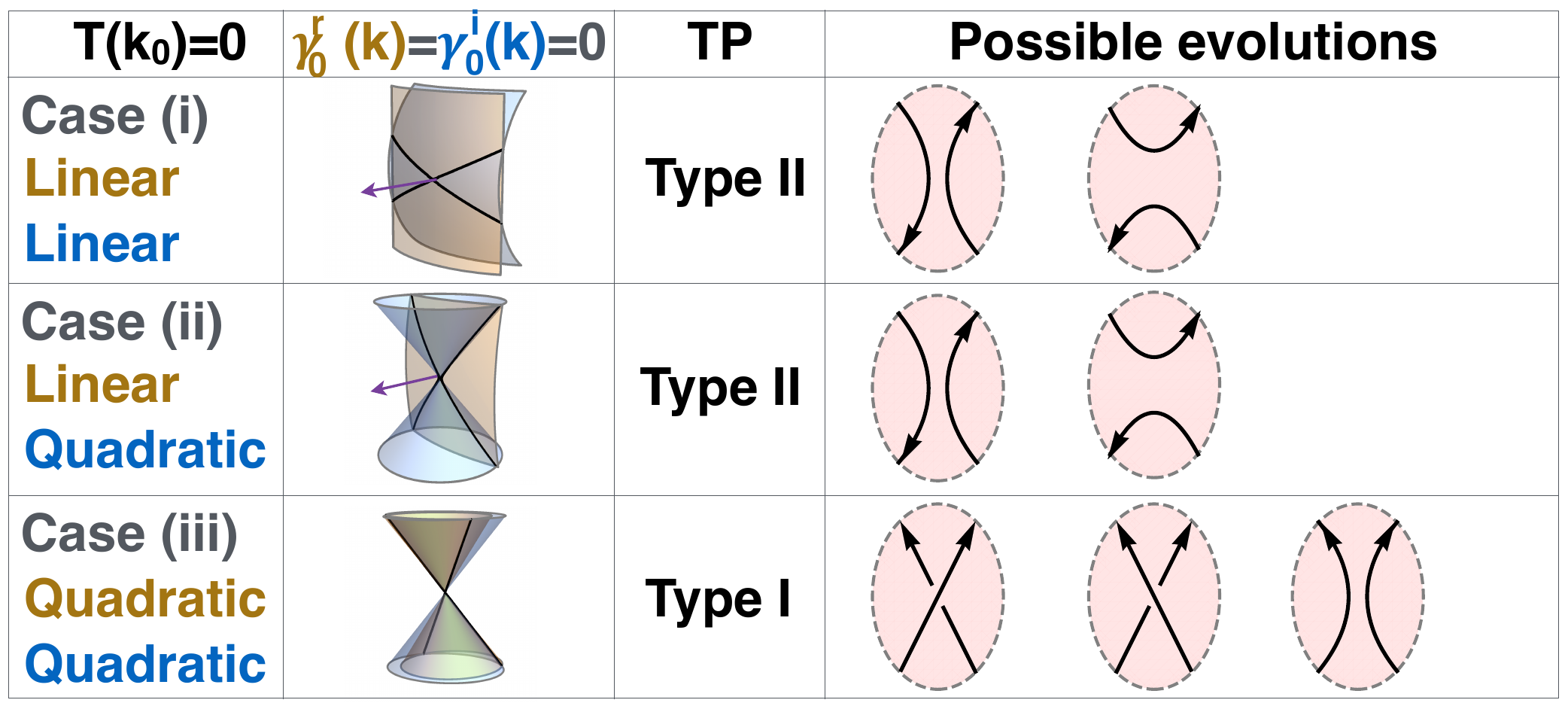}}
\caption{ The classification of the local evolution of two nodal lines with a single TP in the viewpoint of the {\em natural projective plane}, which is the plane spanned by the two tangential vectors at the TP of the two nodal lines. The first column shows the low energy theory around TP can be linear (non-vanishing gradient) or quadratic (vanishing gradient) in different cases. Here linear means linear dispersion of $\gamma_0^r$ or $\gamma_0^i$ along one direction in the BZ. This limits the possible geometry forms of the surface $\gamma_0^{r/i}(\bm{k})=0$ as shown in the second column. The third column shows the possible local evolutions of nodal lines under the constraint low energy theory of TP. 
}\label{F3}
\end{figure}

{\em Physical constraint---} We show that not only the orientations of the lines but also the energy dispersions limit the possibilities of the local evolutions near the TPs. 
To show the limitation from the dispersions, we first study the conditions for the emergence of TPs in the Hamiltonian (\ref{H0}). Mathematically, these TPs are considered as singularity points of the nodal lines \cite{SM1}, 
which are defined by the vanishing of the tangent vector along the nodal lines at the points.  Since the nodal line is located at the intersection of two surfaces $\gamma_0^r(\bm{k})=0$ and $\gamma_0^i(\bm{k})=0$, for a point $\bm{k}_0$ on the nodal line, the tangent vector $\bm{T}(\bm{k}_0)$ is perpendicular to the two normal directions $\vec{\nabla}_{\bm{k}} \gamma_0^r(\bm{k}_0)$ and $\vec{\nabla}_{\bm{k}} \gamma_0^i(\bm{k}_0)$, where $\vec{\nabla}_{\bm{k}}=(\partial_{k_x},\partial_{k_y},\partial_{k_z})$. In this regard, the tangent vector at point $\bm{k}_0$ along the nodal line is given by 
\begin{equation}
\bm{T}(\bm{k}_0)=\vec{\nabla}_{\bm{k}} \gamma_0^r(\bm{k}_0)\times\vec{\nabla}_{\bm{k}} \gamma_0^i(\bm{k}_0). \label{E3}
\end{equation}
Since the TP in the nodal line belongs to singularity point, the momentum $\bm{k}_{\rm{TP}}$ at the TP obeys $\gamma_0^r(\bm{k}_{\rm{TP}})=\gamma_0^i(\bm{k}_{\rm{TP}})=\bm{T}(\bm{k}_{\rm{TP}})=0$. To have $\bm{T}(\bm{k}_{\rm{TP}})=0$, the TP evolution is classified as the three cases: (i) the two gradients are parallel ($\vec{\nabla}_{\bm{k}} \gamma_0^r(\bm{k}_{\rm{TP}})=c\vec{\nabla}_{\bm{k}} \gamma_0^i(\bm{k}_{\rm{TP}})\neq0$), (ii) one of the gradients vanishes, and (iii) both vanish as shown in Fig.~\ref{F3}.  

\begin{figure}[t]
\centerline{\includegraphics[height=3.8cm]{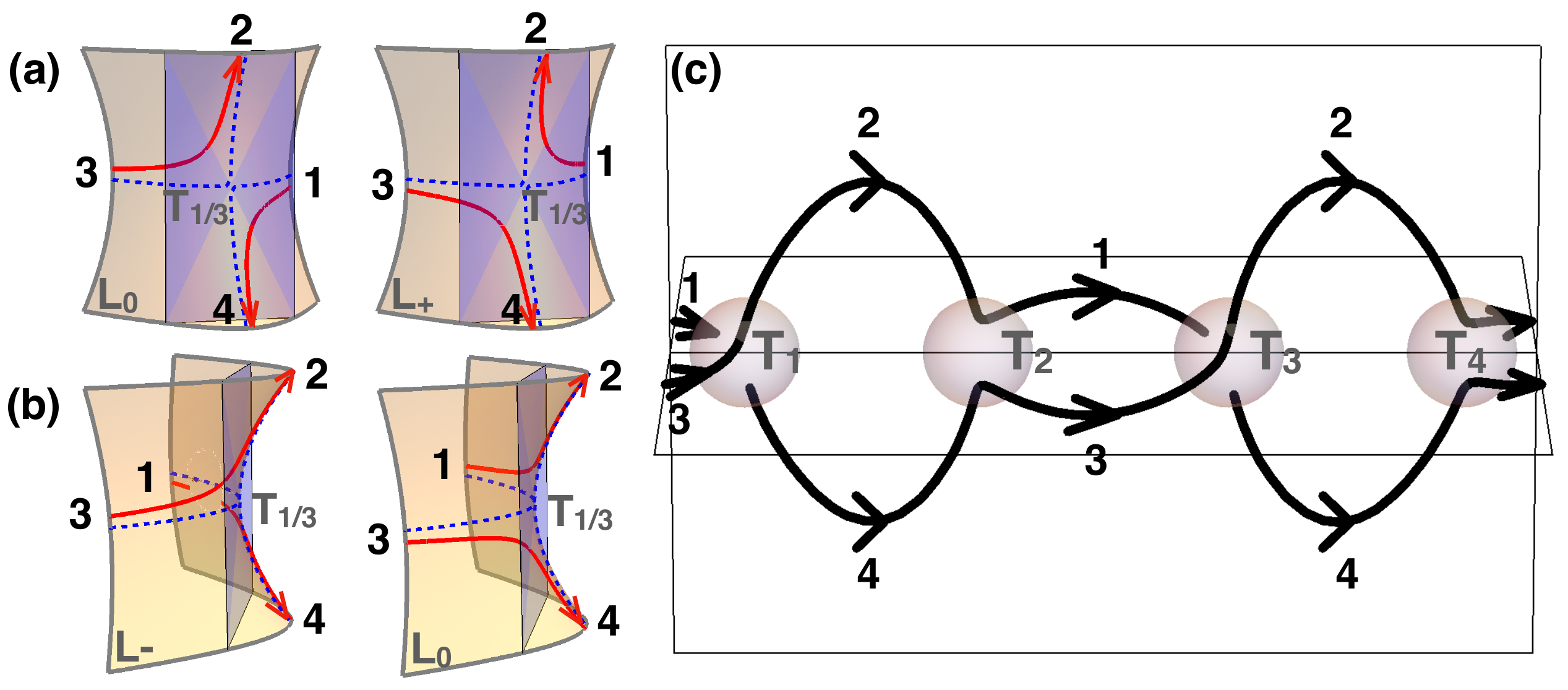}}
\caption{Hopf-link semimetal. (a) shows the physical constraint to the local evolutions of the TPs $T_{1/3}$. Only $L_{0/+}$ of type II are possible in the natural projective plane (blue plane). However, if we rotate the figures in (a), the local evolutions becomes $L_{-/0}$ of type I as shown in (b). (c) shows the Hopf-link semimetal can be obtained from the model in Fig. \ref{F2} (a) by adding the following perturbation $\lambda h_1(\bm{k},\lambda)=i\lambda \sin 2k_x$. 
 \label{deform_TP} }
\end{figure}

In particular, for case (i) and (ii), at least one of the two surfaces must have nonzero gradients at $\bm{k}_{\rm{TP}}$, which satisfies $\vec{\nabla}_{\bm{k}} \det[h_0(\bm{k}_{\rm{TP}})]\neq0$ leading to linear dispersions. Since the types of local evolution depends on the choice of projective plane, we fix a special projective plane spanned by the two tangential vectors at the TP of the two nodal lines, which is dubbed as {\em natural projective plane}. For case (i) and (ii), the natural projective plane is perpendicular to the normal vector $\vec{\nabla}_{\bm{k}} \gamma_0^{r/i}(\bm{k}_{\rm{TP}})$. It can be shown that for the first two cases the TP must be type II in the natural projective plane~\cite{SM1}, and the local evolution near the TP is limited to the two possibilities shown in the first two rows of Fig.~\ref{F3}. The reason is that the surface with nonzero gradient at the TP (yellow surface) can always be mapped to the natural projective plane (light blue plane) as shown in Fig.~\ref{deform_TP}(a). This forbids the emergence of $L_-$ in Fig.~\ref{F1} (b2). 
Here we emphasize that the limitation of the local evolution holds only in the {\em natural projective planes}. Choosing another projective plane, we have to transfer the constraint from the natural projective plane to the chosen plane. For example, Fig.~\ref{deform_TP}(a) shows only $L_+$ and $L_0$ of type II are the only two possible local evolutions in the natural projective plane. By changing a different point of view and applying this evolution constraint, Fig.~\ref{deform_TP}(b) in the new projective plane shows that the two possible local evolutions becomes $L_-$ and $L_0$ of type I in Fig.~\ref{F1} (b1). 

In contrast to the former two cases, in case (iii) neither $\gamma_0^r(\bm{k})$ nor $\gamma_0^i(\bm{k})$ possesses linear terms near $\bm{k}_{\rm{TP}}$ as shown in the third row of Fig.~\ref{F3}. By assuming quadratic terms of $\bm{k}$,  in the {\em natural projective plane}, the line arrangement at the TP is constrained to type I, and there are three possible evolutions as shown in Fig.~\ref{F3}~\cite{SM1}. 

{\em Nodal chain semimetals---} To demonstrate the tools we established  for the study of the knot evolution in semimetals, we are back to the nodal chain semimetal with 4 TPs protected by chiral and two mirror symmetries~\cite{reveiw_semimetal_b,Yan:2018aa}. The linear terms of the four TPs are nonzero in the direction perpendicular to the two mirror planes. We consider a specific nodal chain described by the chiral symmetric Hamiltonian (\ref{H0}) with $h_0(\bm{k})=2\cos 2k_x+\cos k_x+3\cos k_y-3\cos k_z-1/10-2i \sin k_y\sin k_z$ as shown in Fig.~\ref{F2}(a). Although there are two spatially separated nodal chains in this model~\cite{SM1}, we focus on one in the two mirror planes ($k_y=0$ and $k_z=0$) and the chain is marked by $L_{n_1n_2n_3n_4}$ representing its topology is determined by the evolution of the four TPs $T_i$. The linear dispersion near the TP leads to the non-zero gradient $\vec{\nabla}_{\bm{k}} \det [ h_0(\bm{k}_{\rm{TP}})]\neq 0$. First, consider the local evolutions at $T_1,\ T_3$, which can have only $L_+$ and $L_0$ of type II in the natural projective planes (blue planes) as shown in Fig.~\ref{deform_TP}(a). In the other view angle for the knot diagrams in Fig.~\ref{F2}(b-d), $L_-$ and $L_0$ of type I are only two possible evolutions as shown in Fig.~\ref{deform_TP}(b). Similarly, the TPs $T_2, T_4$ in the projective planes can evolve only to $L_+$ and $L_0$ of type I in the knot diagram. As a result, the generated phases are constrained to be $L_{n_1m_1n_2m_2}$, where $n_i=0,1$ and $m_i=0,-1$. Due to this constraint from the linear dispersion, globally the chain can evolve to an unknot ($|n|=1$), an unlink ($|n|=0$), or a Hopf-link ($|n|=2$) listed in Table \ref{T1} with red color. As shown in Fig.~\ref{deform_TP} (c), we have a nodal  Hopf-link under the perturbation Eq.~\ref{E2} with $\lambda h_1(\bm{k},\lambda)=i\lambda \sin 2k_x$. Physically, we have two ways to control perturbations. One is to add the pressure of a material that breaks the mirror symmetry. The other way is that in the photonic lattice, the lattice can be designed artificially. Hence the mirror symmetry breaking term can be added in a controlled way~\cite{Lu622}. Finally, using the same recipe, we can show that the nodal chain semimetal ($L_{n_1m_1}$) with 2 TPs cannot evolve to a Hopf-link ($L_{\pm \pm}$) when $\vec{\nabla}_{\bm{k}} \det [ h_0(\bm{k}_{\rm{TP}})]\neq 0$ at each TP~\cite{SM1}.

{\em Non-Hermitian exceptional line semimetals---} This recipe studying the knot topology can even be extended to the non-Hermitian system~\cite{benderMakingSenseNonHermitian2007b,fengNonHermitianPhotonicsBased2017b,el-ganainyNonHermitianPhysicsPT2018d,ozdemirParityTimeSymmetry2019a,miriExceptionalPointsOptics2019a,shenTopologicalBandTheory2018d,yaoEdgeStatesTopological2018b,yaoNonHermitianChernBands2018b,yokomizoNonBlochBandTheory2019a,zhangCorrespondenceWindingNumbers2019},
namely, the 3D non-Hermitian exceptional line semimetals ~\cite{PhysRevLett.118.045701,PhysRevB.97.075128,PhysRevA.98.042114,PhysRevB.99.081102,PhysRevB.99.161115,PhysRevB.99.041116,PhysRevB.99.161115,wangNonHermitianNodallineSemimetals2019b,PhysRevLett.123.066405}. While in Hermitian systems the nodal lines require symmetry protection~\cite{reveiw_semimetal_b}, the non-Hermitian exceptional lines are robust against any small perturbation even in the absence of any symmetries~\cite{shenTopologicalBandTheory2018d,PhysRevB.99.081102}. We here focus on a general $N$-band non-Hermitian tight-binding Hamiltonian $\mathcal{H}_{nH}(\bm{k})$ with periodic boundary condition or with no skin modes~\cite{yaoEdgeStatesTopological2018b,yaoNonHermitianChernBands2018b,yokomizoNonBlochBandTheory2019a,zhangCorrespondenceWindingNumbers2019}. According to the characteristic polynomial of the Hamiltonian 
\begin{equation}
f(E,\bm{k})=\det[E-\mathcal{H}_{nH}(\bm{k})]=\Pi_{i=1}^N[E-E_i(\bm{k})],
\end{equation} 
the condition for the emergence of band degeneracy $E_i(\bm{k})=E_j(\bm{k})$ requires $f(E,\bm{k})=\partial_E f(E,\bm{k})=0$, which is equivalent to~\cite{SM1}
\begin{equation}
\Delta_{f}(\bm{k})=\prod_{i<j} [E_i(\bm{k})-E_j(\bm{k})]^2=0,
\label{E8}
\end{equation}
where $\Delta_{f}(\bm{k})$ is the discriminant of the characteristic polynomial $f(E,\bm{k})$ as a function of $E$~\cite{SM1,Discriminant}. For example, if $f(E)=aE^2+bE+c$, $\Delta_f=b^2-4ac$.
Hence, the solution of Eq.~\ref{E8} must be a set of 1D degeneracy lines in the 3D BZ. Using the Sylvester matrix of the characteristic polynomial to build the discriminant~\cite{SM1, Resultant1,Resultant2}, we can show the discriminant is a single-valued function of $\bm{k}$. Therefore, the topological charge can be defined by the quantized winding number in Eq.~\ref{WN} with $h_0(\bm{k})=\Delta_{f}(\bm{k})$, and the non-zero winding number protects the degeneracy line and determines the knot orientation. In particular, in presence of arbitratry perturbations, the degeneracies are called {\em stable exceptional lines}~\cite{PhysRevLett.123.066405,2019arXiv191202788Y}, where the non-Hermitian Hamiltonian is not diagonalizable~\cite{miriExceptionalPointsOptics2019a}. Since the mathematical structures of the Hermitian chiral symmetric systems and non-Hermitian ones are identical, we follow the same recipe above to characterize the evolution of the degeneracy lines near TPs in the non-Hermitian systems with the identical constraint of the local evolution at each TP.    

In summary, topologically-protected lines emerge in several distinct condensed matter systems, such as Hermitian chiral-symmetric semimetals and non-Hermitian systems. We start with nodal lines with several TPs; the Jones polynomial characterizes the knot topology of the lines with orientation in 3D BZ, and the topology essentially is only determined by the local evolution near each TP. The low energy theory limits the line orientation at any TP; furthermore, if $\vec{\nabla}_{\bm{k}} \det [ h_0(\bm{k}_{\rm{TP}})]$ or $\vec{\nabla}_{\bm{k}}\Delta_f(\bm{k}_{\rm{TP}})$ does not vanish at the TP,  
the corresponding local evolution is limited to two possible ones. Using the nodal chain semimetal with 4 TPs as an example, we can show how can we calculate the Jones polynomial for the generated phases. Our methodology provides general rules to the evolution of the topologically-protected lines with TPs and paves the way toward searching for exotic topological knot-node semimetals. 

{\em Acknowledgements}--- The work is supported by the Ministry of Science and Technology of China 973 program (No. 2015CB921300, No.~2017YFA0303100), National Science Foundation of China (Grant No.  NSFC-11888101, 1190020, 11534014, 11334012), and the Strategic Priority Research Program of CAS (Grant No.XDB07000000). C.-K.C. is supported by the Strategic Priority Research Program of the Chinese Academy of Sciences (Grant XDB28000000). 

\bibliographystyle{apsrev4-1}
\bibliography{TOPO_v14}

\onecolumngrid
\newpage


\section*{Supplementary material}

# Supplementary Material for
# "Jones polynomial and knot transitions in topological semimetals"

Zhesen Yang,[1,2] Ching-Kai Chiu,[3,*] Chen Fang,[1,†] and Jiangping Hu[1,3,4,‡]

[1]*Beijing National Laboratory for Condensed Matter Physics,
and Institute of Physics, Chinese Academy of Sciences, Beijing 100190, China*
[2]*University of Chinese Academy of Sciences, Beijing 100049, China*
[3]*Kavli Institute of Theoretical Sciences, University of Chinese Academy of Sciences, Beijing, 100190, China*
[4]*Collaborative Innovation Center of Quantum Matter, Beijing, China*
(Dated: December 11, 2019)

The contents of the Supplementary Material are summarized as follows: (i) derivation of the winding number of nodal lines protected by chiral symmetry (section I); (ii) introduction of knot theory and Jones polynomial (section II); (iii) classification of TPs and local evolutions (section III, IV); (iv) extended skien relations (section V); (v) introduction of singularity points (section VI); (vi) physical constraint of local evolutions (section VII); (vii) nodal chain semimetals protected by two mirror symmetries (section VIII, IX); (viii) introduction of discriminant (section X).

## I. TOPOLOGICAL CHARGE OF NODAL LINES PROTECTED BY CHIRAL SYMMETRY

The topological charge of nodal lines protected by chiral symmetry is the winding number, which can be defined from the $Q$ matrix [1],

$$Q(\boldsymbol{k}) = \sum_{0 < E_\alpha}^{N_+} |\Psi^\alpha(\boldsymbol{k})\rangle\langle\Psi^\alpha(\boldsymbol{k})| - \sum_{0 > E_\beta}^{N_-} |\Psi^\beta(\boldsymbol{k})\rangle\langle\Psi^\beta(\boldsymbol{k})|, \tag{1}$$

where $|\Psi^\alpha(\boldsymbol{k})\rangle$ is the eigenstate of $\mathcal{H}(\boldsymbol{k}) = h(\boldsymbol{k})\tau_+ + h^\dagger(\boldsymbol{k})\tau_-$ with eigenenergy $E_\alpha(\boldsymbol{k})$. Since the system has chiral symmetry, $Q$ matrix can be written as

$$Q(\boldsymbol{k}) = q(\boldsymbol{k})\tau_+ + q^\dagger(\boldsymbol{k})\tau_-, \tag{2}$$

where $q(\boldsymbol{k})$ is an unitary $N \times N$ matrix. It is easy to show that $\mathcal{H}(\boldsymbol{k})$ commutes with $Q(\boldsymbol{k})$ since

$$\mathcal{H}(\boldsymbol{k})Q(\boldsymbol{k}) = \sum_{0 < E_\alpha}^{N_+} E_\alpha(\boldsymbol{k})|\Psi^\alpha(\boldsymbol{k})\rangle\langle\Psi^\alpha(\boldsymbol{k})| - \sum_{0 > E_\beta}^{N_-} E_\beta(\boldsymbol{k})|\Psi^\beta(\boldsymbol{k})\rangle\langle\Psi^\beta(\boldsymbol{k})| = Q(\boldsymbol{k})\mathcal{H}(\boldsymbol{k}). \tag{3}$$

Rewritten the commutation relation

$$0 = [\mathcal{H}(\boldsymbol{k}), Q(\boldsymbol{k})] = \begin{pmatrix} h(\boldsymbol{k})q^\dagger(\boldsymbol{k}) - q(\boldsymbol{k})h^\dagger(\boldsymbol{k}) & 0 \\ 0 & h^\dagger(\boldsymbol{k})q(\boldsymbol{k}) - q^\dagger(\boldsymbol{k})h(\boldsymbol{k}) \end{pmatrix}, \tag{4}$$

one can obtain

$$\det\left[h(\boldsymbol{k})q^\dagger(\boldsymbol{k})\right] = \det\left[q(\boldsymbol{k})h^\dagger(\boldsymbol{k})\right] = \det\left[\left(h(\boldsymbol{k})q^\dagger(\boldsymbol{k})\right)^\dagger\right] = \det\left[h(\boldsymbol{k})q^\dagger(\boldsymbol{k})\right]^* \tag{5}$$

The above equation implies that $\det[h(\boldsymbol{k})q^\dagger(\boldsymbol{k})]$ is a real function of $\boldsymbol{k}$ so that $\ln\det[h(\boldsymbol{k})] + \ln\det[q^*(\boldsymbol{k})] = r(\boldsymbol{k})$ is a single-valued function, which leads to $\oint d\boldsymbol{k} \cdot \nabla_{\boldsymbol{k}} r(\boldsymbol{k}) = 0$. Using the property of the unitary matrix $q(\boldsymbol{k})$, we have $\ln\det[q^*] +$

---

*Electronic address: qiujingkai@ucas.edu.cn
†Electronic address: cfang@iphy.ac.cn
‡Electronic address: jphu@iphy.ac.cn

$\ln \det[q] = 0$. Hence, the winding number with the closed integral path $\Gamma(\boldsymbol{K}_0)$ can be directly written in form of $h(\boldsymbol{k})$

$$
\begin{aligned}
\nu &= \frac{i}{2\pi} \oint_{\Gamma(\boldsymbol{K}_0)} d\boldsymbol{k} \cdot \mathrm{Tr}\left[q^{-1}(\boldsymbol{k})\nabla_{\boldsymbol{k}}q(\boldsymbol{k})\right] \\
&= \frac{i}{2\pi} \oint_{\Gamma(\boldsymbol{K}_0)} d\boldsymbol{k} \cdot \nabla_{\boldsymbol{k}} \mathrm{Tr}[\ln q(\boldsymbol{k})] \\
&= \frac{i}{2\pi} \oint_{\Gamma(\boldsymbol{K}_0)} d\boldsymbol{k} \cdot \nabla_{\boldsymbol{k}}(\ln \det[q(\boldsymbol{k})]) \\
&= \frac{i}{2\pi} \oint_{\Gamma(\boldsymbol{K}_0)} d\boldsymbol{k} \cdot \nabla_{\boldsymbol{k}}(\ln \det[h(\boldsymbol{k})]).
\end{aligned}
\tag{6}
$$

This means the topological charge of the nodal line is determined by the winding number of $\det[h(\boldsymbol{k})] = \gamma_0^r(\boldsymbol{k}) + i\gamma_0^i(\boldsymbol{k})$. It should be emphasized that the sign of the winding number is determined by the orientation of the integral path $\Gamma(\boldsymbol{K}_0)$. If we reverse the integral path, the winding number will change its sign. Thus one can always find a suitable path orientation that the winding number is positive, if the winding number is nonzero. Finally, one can assign a direction to the point $\boldsymbol{K}_0$ on the nodal line, which is the normal vector of the counterclockwise integral path centered at $\boldsymbol{K}_0$ corresponding to the positive winding number. By marking an arrow pointing this normal direction along the line, one can assign an orientation to these nodal lines. In particular, for winding number $\nu = \pm 1$, the normal direction of the nodal line is aligned with $\vec{\nabla}_{\boldsymbol{k}}\gamma_0^i(\boldsymbol{K}_0) \times \vec{\nabla}_{\boldsymbol{k}}\gamma_0^r(\boldsymbol{K}_0)$. As we introduce a plane to interact the nodal line, the sign of the winding number of the intersection point is given by

$$
\mathrm{sign}(\nu_{\vec{n}}) = \vec{n} \cdot \vec{\nabla}_{\boldsymbol{k}}\gamma_0^i(\boldsymbol{K}_0) \times \vec{\nabla}_{\boldsymbol{k}}\gamma_0^r(\boldsymbol{K}_0),
\tag{7}
$$

where $\vec{n}$ is the normal vector.

## II. BRIEF INTRODUCTION OF KNOT THEORY

### A. Basic concepts

Mathematically, a knot $K$ is an embedding of a circle in three dimensional (3D) space. This 3D space can be $\mathbb{R}^3$, $S^3$, $\mathbb{T}^3$ (the 3D Brillouin zone) and so on. One of simplest examples of nontrivial knot is the trefoil knot, as shown in Fig. 1 (a). A link $L$ is collection of several knots, which do not intersect but may be linked (or knotted) together. It can be written as $L = K_1 \cup \cdots \cup K_n$, where $K_i$ is the $i$th component of the link. The simplest example of nontrivial link is the Hopf-link, which is shown in Fig. 1 (d). By definition, one can realize that the knot is a special type of link, which only has one component. Notice that the definition of knot and link forbids the intersections, namely they can not have any touching point (TP). Two links $L$ and $L'$ are equivalent, written as $L \simeq L'$, if one can take $L$ and deform it in the 3D space to obtain $L'$ by non-broken bending and stretching without tearing the lines apart.

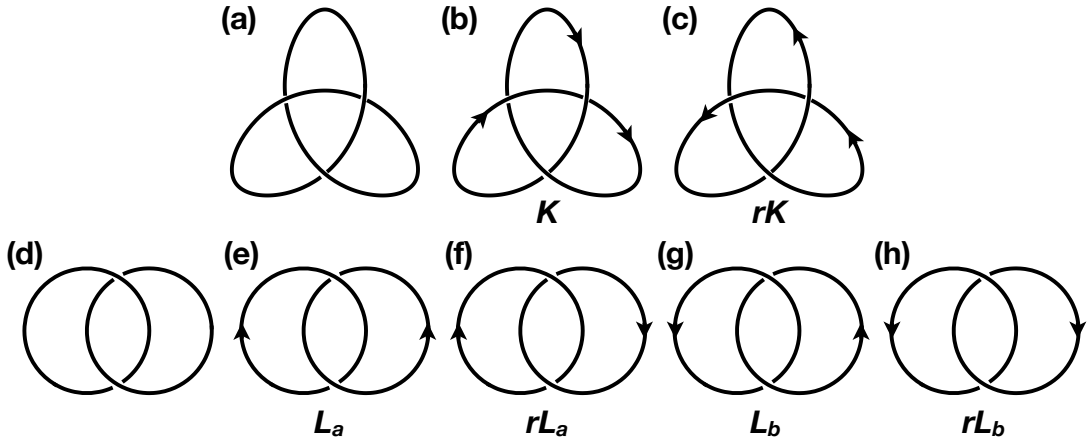

FIG. 1: Trefoil knot and Hopf-link. (a) shows the diagram of trefoil knot, which has three crossings. (b) and (c) show two different kinds of orientations of trefoil knot, labeled by $K$ and $rK$. Here $rK$ is the reverse of $K$. (d) shows the diagram of Hopf-link. (e)-(h) show its orientations.

Although the knots or links are defined in 3D space, they can also be represented in a two dimensional way by using knot and link diagrams. A diagram of a link $L$ is a projection of $L$, together with the data of over- and under-crossings. For example, Fig. 1 (a) and (d) are the diagrams of the trefoil knot and Hopf-link. They have three and two crossings respectively. Obviously, depending on the choice of projective plane, the link diagrams can be different, but they should represent the same link. It is useful to introduce the orientation of a link, which is defined by a choice direction to each component in a consistent way and an oriented link is a link together with an orientation. Obviously, a link can be assigned to different orientations. The reverse of a link $L$ is defined by reversing all the orientations for each components, which is labeled by $rL$. For example, the trefoil knot has two orientations as shown in Fig. 1 (b) and (c). They are related by reverse. The Hopf-link has four possible orientations, which are shown in Fig. 1 (e)-(h).

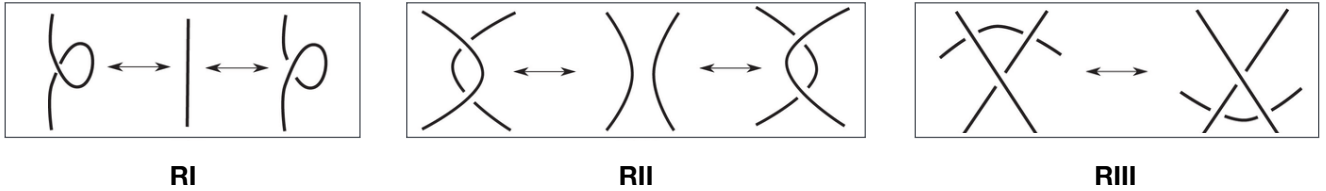

FIG. 2: Reidemeister moves. There are three types of Reidemeister moves, which are labeled by RI, RII and RIII. If two diagrams are related by a sequence of Reidemeister moves, then they represent equivalent knots or links

For the link diagrams, the topological equivalent classes can be defined by the *Reidemeister moves* as shown in Fig. 2, that is if two diagrams are related by a sequence of Reidemeister moves, then they represent equivalent knots or links. For the oriented links, one can also define the equivalence by the Reidemeister moves. Since the oriented trefoil knot $K$ cannot deform to another trefoil knot $rK$ through any Reidemeister moves, $K$ and $rK$ are inequivalent. On the other hand, Reidemeister moves can connect the two Hopf-links $L_a$ and $L_b$; hence, the oriented links are equivalent.

### B. Knot invariants

#### 1. Linking number

For the oriented links, one can define the sign for every crossing of two disconnected lines, as shown in Fig. 3. Based on the

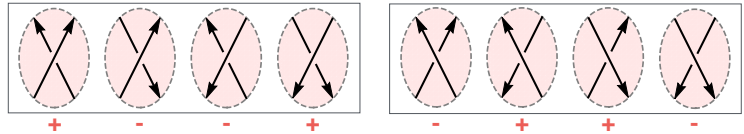

FIG. 3: The sign of oriented crossings.

sign of the crossings, one can define the linking number of an oriented link $L$, which is the one half of the summation of the signs of all the crossings between different components of $L$. For example, if $L$ only have two components $\alpha$ and $\beta$, then the linking number is

$$lk(\alpha, \beta) = \frac{1}{2} \sum_{p \in \alpha \cup \beta} \text{sign}(p) \qquad (8)$$

where $\text{sign}(p)$ is the sign of the crossings between $\alpha$ and $\beta$ component. For example, in Fig. 1 (e)-(h), their linking numbers are $lk(L_a) = lk(rL_a) = -1$, $lk(L_b) = lk(rL_b) = 1$. Notice the linking number is invariant on the choice of diagrams and local deformations, which means it is a knot invariant. The proof can be done by showing that the linking number is invariant under Reidemeister moves [2, 3].

## 2. Jones polynomial

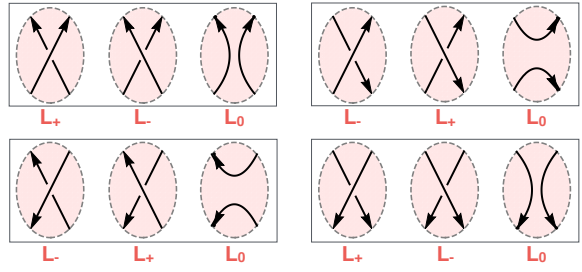

FIG. 4: Skein relation with different orientations. The $L_\pm$ is determined by the *sign* of the crossing.

The Jones polynomial $J$ is a topological invariant for oriented links with the following properties:

- it takes values in $\mathbb{Z}\left[t^{1/2}, t^{-1/2}\right]$;

- it satisfies $J(O) = 1$, where $O$ denotes the unknot.

- it satisfies the skein relation

$$t^{-1} J\left(L_+\right) - t J\left(L_-\right) + \left(t^{-1/2} - t^{1/2}\right) J\left(L_0\right) = 0 \tag{9}$$

whenever $L_+$, $L_-$ and $L_0$ are three oriented links, which have diagrams that are identical except in a small region where they are given as shown in Fig. 4.

The skein relation plays the central role of knot polynomial. Now we will show some examples of the calculation of Jones polynomial based on the skein relation.

The first example we considered is shown in Fig. 5. Obviously, $L_{+000}$ and $L_{-000}$ are both equivalent to an unknot. Hence $J(L_{+000}) = J(L_{-000}) = 1$. Using the skein relation, one can obtain,

$$t^{-1} - t + \left(t^{-1/2} - t^{1/2}\right) J\left(L_{0000}\right) = 0. \tag{10}$$

After some simple calculation, we have $J(L_{0000}) = -t^{1/2} - t^{-1/2}$. Hence, the Jones polynomials of single unknot is different from the one of the two separated unknots.

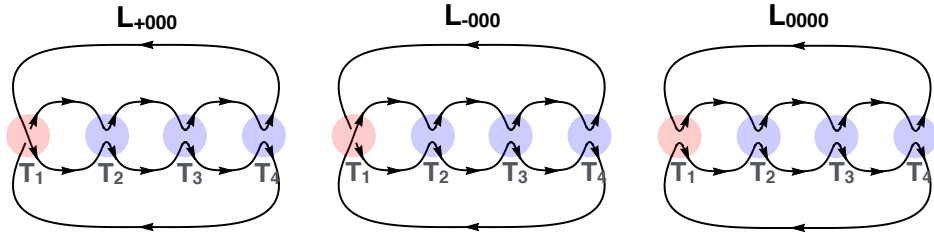

FIG. 5: Example 1

The second example is shown in Fig. 6. The skein relation relates the Hopf-link to the unknot. Since we know $J(L_{+00-}) = -t^{1/2} - t^{-1/2}$ and $J(L_{+000}) = 1$. Hence based on the skein relation,

$$t^{-1} J(L_{+00+}) - t \left(-t^{1/2} - t^{-1/2}\right) + \left(t^{-1/2} - t^{1/2}\right) = 0. \tag{11}$$

One can obtain, the Hopf-link has $J(L_{+00+}) = -t^{5/2} - t^{1/2}$.

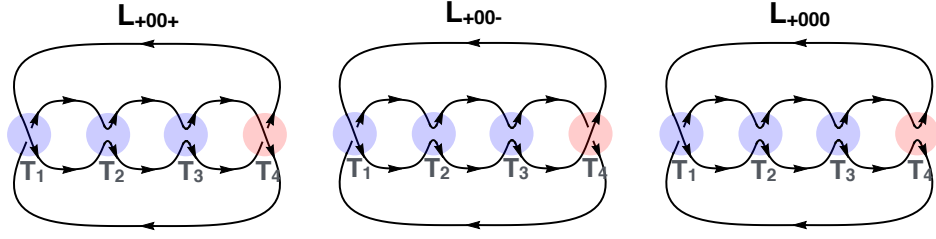

FIG. 6: Example 2

## III. CLASSIFICATION OF TPS

In the main text, we have mentioned that there exist two types of TPs, namely type I and type II. Here we show this. Without considering the orientations, a TP with two nodal lines has four arcs. If we assign an orientation to each arcs, there exist $2^4$ types of orientations. We use $i$ ($\underline{i}$) to label the $i$th arc pointing at (pointing out) the TP. Fig. 7 (a) shows some examples. Now we show

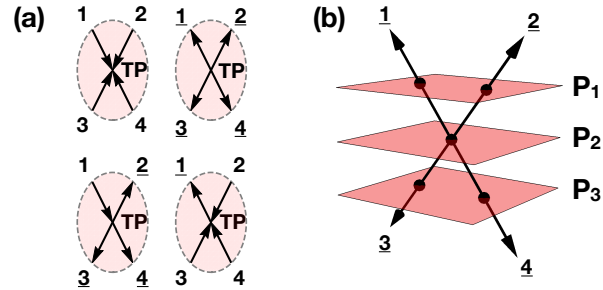

FIG. 7: The inconsistent orientations of the TP. (a) shows some example. (b) shows the proof.

that the following orientated TPs are impossible,

$$1234, \underline{1}234, 1\underline{2}34, 12\underline{3}4, 123\underline{4}, 1\underline{2}3\underline{4}, \underline{1}2\underline{3}4, \underline{1}\underline{2}3\underline{4}, \underline{1}2\underline{3}\underline{4}, 1\underline{2}\underline{3}\underline{4}. \tag{12}$$

Here we only prove the case $12\underline{3}4$. The proof for the other cases are straightforward. As shown in Fig. 7(b), in the plane $P_1$, the system has two nodal points, whose winding numbers are $+1$. Thus under the evolution of the 2D plane, the two nodal points can touch together, as shown in Fig. 7 (b) with $P_2$. Since the topological charge of the two nodal points can't change, the winding number of the TP in the plane $P_2$ must $+2$. However, in the plane $P_3$, the charge of the two nodal points are $-1$, which implies the charge of the TP is $-2$. This is inconsistent with the above result. Thus this kind of orientation is impossible.

Finally, we obtain the following kinds of orientations, $1\underline{2}\underline{3}\underline{4},\underline{1}2\underline{3}4, \underline{1}2\underline{3}4, 12\underline{3}\underline{4}, 1\underline{2}3\underline{4}$ and $\underline{1}2\underline{3}\underline{4}$, which are shown in Fig. 8. We

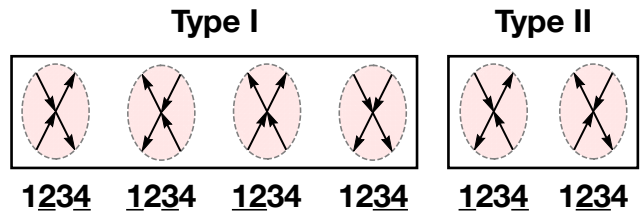

FIG. 8: The consistent orientations of the TP. , which can be classified by type I and type II.

will study their corresponding local evolutions.

## IV. CLASSIFICATION OF LOCAL EVOLUTIONS

In the main text, we have mentioned that all the perturbation generated phases can be labeled by the local evolutions around every TP. Hence a complete classification of local evolutions is necessary. As discussed in the previous section, only the orientations shown in Fig. 8 are possible. Since 1̲2̲3̲4̲, 1̲2̲3̲4̲ and 1̲2̲3̲4̲ can be obtained from 1̲2̲3̲4̲ through $\pi/2$ or $\pi$ or $3\pi/2$ rotations, and 1̲2̲3̲4̲ can be obtained from 1̲2̲3̲4̲ through $\pi$ rotation, we only focus on 1̲2̲3̲4̲ and 1̲2̲3̲4̲. Fig. 9 shows their corresponding possible local evolutions. Having obtained the mathematical local evolutions, we now show the skein relation

### Type I TP and local evolutions

### Type II TP and local evolutions

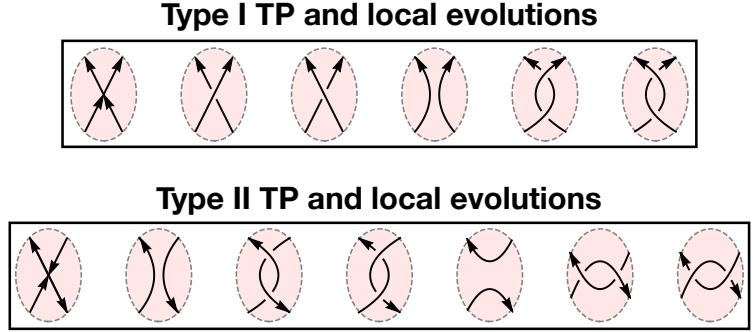

FIG. 9: The mathematically allowed local evolutions around different kinds of TPs.

between them.

## V. EXTENSION OF SKIEN RELATION

In this section, we will label the local evolutions based on the skein relation of the Jones polynomials. Fig. 10 shows all the extended skein relation of type I/II TPs and the corresponding local evolutions. Having labeled all the kinds of local evolutions, one can represent all the perturbation generated phases from the critical phase, for example $L_{n_1,\dots,n_m}^{\alpha_1,\dots,\alpha_m}$, where $\alpha_i = a, b, c$ or $a, b, c, d$ based on the types of TPs. However, for many simple cases (where the TP has nonzero quadratic terms), $L_-^b$ and $L_+^c$ are impossible to occur in the natural projective plane. Thus, in order to make the discussion more clear, we only provide the local evolutions shown in Fig. 1 in the main text.

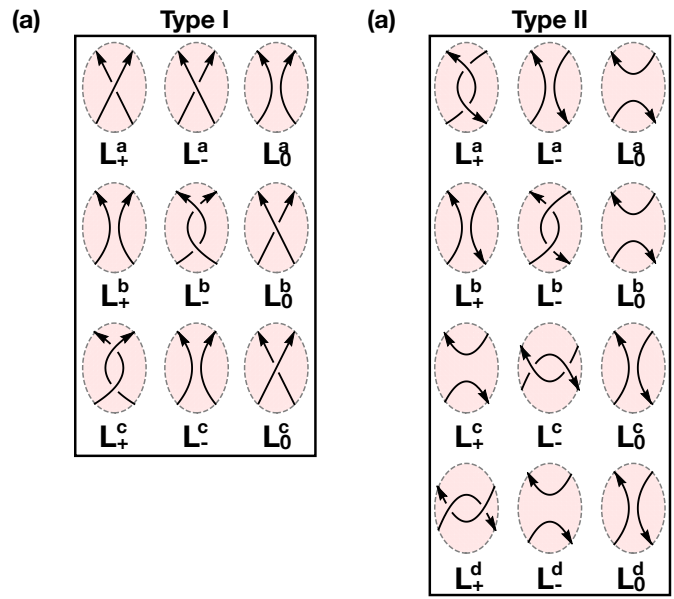

FIG. 10: All the extended skein relation of type I/II TPs and the corresponding local evolutions.

Here only $L_{+,-,0}^a$ in the type I evolution are the standard skein relation of the Jones polynomials. The proof of all the other

extended skein relation can be shown based on the standard skein relation. Here we only prove one example. The proof of the other cases are straightforward. The proof is shown in Fig. 11, which is based on the Reidemeister moves.

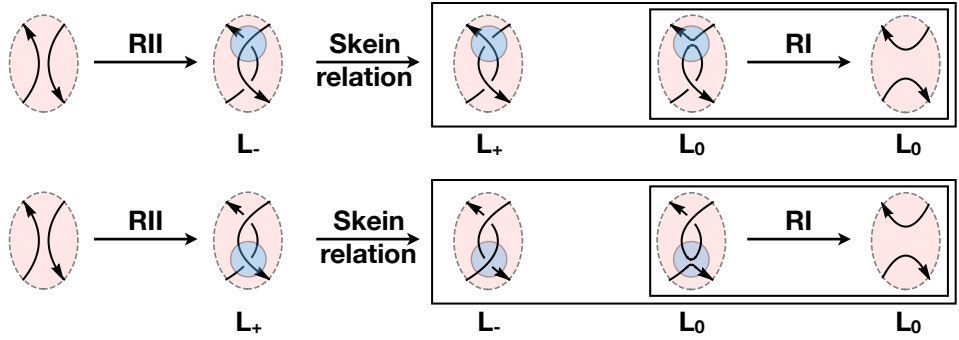

FIG. 11: Proof of the extended skein relation, where RI and RII represent the type I and type II Reidemeister moves respectively.

## VI. SINGULARITY POINT OF A CURVE

As mentioned in the main text, the TP, where two local nodal lines touch together, belongs to the singularity point of the nodal lines. Consider an plane algebraic curve $C$, which is defined by the following polynomial equation,

$$P(x, y) = 0. \tag{13}$$

If a point on $C$, say $(a, b) \in C$, satisfies

$$\frac{\partial P}{\partial x}(a, b) = \frac{\partial P}{\partial y}(a, b) = 0, \tag{14}$$

it is called singularity point of the plane algebraic curve $C$. For example, as shown in Fig. 12, the curve $y^2 - x^3 - x^2 - 1 = 0$ has no singularities, whereas the curves $y^2 - x^3 - x^2 = 0$ and $y^2 - x^3 = 0$ have singularities at the origin.

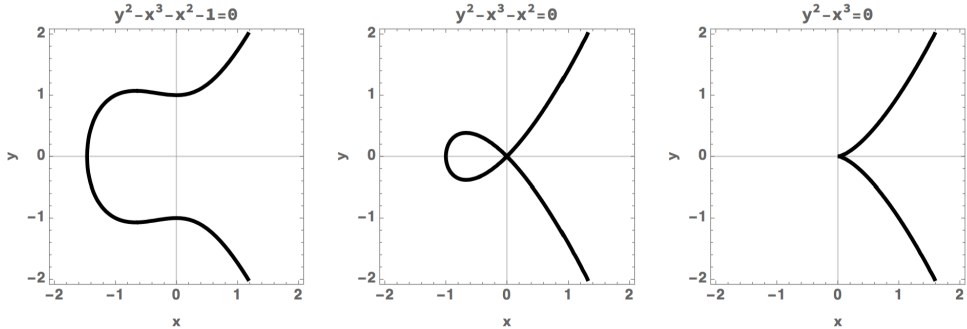

FIG. 12: Plane algebraic curves and singularities. $y^2 - x^3 - x^2 = 0$ and $y^2 - x^3 = 0$ have singularity at the origin.

In the case of nodal line semimetals protected by chiral symmetry, the Hamiltonian can be written as $\mathcal{H}_0(\boldsymbol{k}) = h_0(\boldsymbol{k})\sigma_+ + h_0^\dagger(\boldsymbol{k})\sigma_-$. The nodal lines are one dimensional curves in $3d$ BZ, which are determined by the following two equations,

$$\mathrm{Re}\det[h_0(\boldsymbol{k})] = \gamma_0^r(\boldsymbol{k}) = 0, \qquad \mathrm{Im}\det[h_0(\boldsymbol{k})] = \gamma_0^i(\boldsymbol{k}) = 0. \tag{15}$$

Hence the definition of of singularity point $\boldsymbol{k}_0$ on the nodal line is generalized by the vanishing of the tangent vector, $\boldsymbol{T}(\boldsymbol{k}_0) = 0$. The tangent vector is perpendicular to the normal directions of two surfaces $\mathrm{Re}\det[h_0(\boldsymbol{k})] = 0$ and $\mathrm{Im}\det[h_0(\boldsymbol{k})] = 0$, that is,

$$\boldsymbol{T}(\boldsymbol{k}_0) = \vec{\nabla}_{\boldsymbol{k}}\gamma_0^r(\boldsymbol{k}_0) \times \vec{\nabla}_{\boldsymbol{k}}\gamma_0^i(\boldsymbol{k}_0). \tag{16}$$

## A. Using singularity to determine the linking transition point

Now we will use the condition of singularity to determine the phase transition between a Hopf-link and two separated nodal lines. The model of Hopf-link semimetal can be written $\mathcal{H}(\boldsymbol{k}) = h_x(\boldsymbol{k})\sigma_x + h_z(\boldsymbol{k})\sigma_z$

$$
\begin{aligned}
h_x(\boldsymbol{k}) &= (n_1 - n_3)(n_3 - \lambda n_0) - n_4(n_2 - n_4), \\
h_z(\boldsymbol{k}) &= (n_2 - n_4)(n_3 - \lambda n_0) + n_4(n_1 - n_3),
\end{aligned}
\tag{17}
$$

where $n_0 = k^2 + 1$, $n_1 = 2k_x$, $n_2 = 2k_y$, $n_3 = 2k_z$, $n_4 = k^2 - 1$ and $\lambda$ is the external parameter. Fig. 13 shows the evolution of nodal lines with different values of $\lambda$. One can notice that the Hopf-link nodal line semimetals can emerge in the parameter region $|\lambda| < 1/\sqrt{2}$.

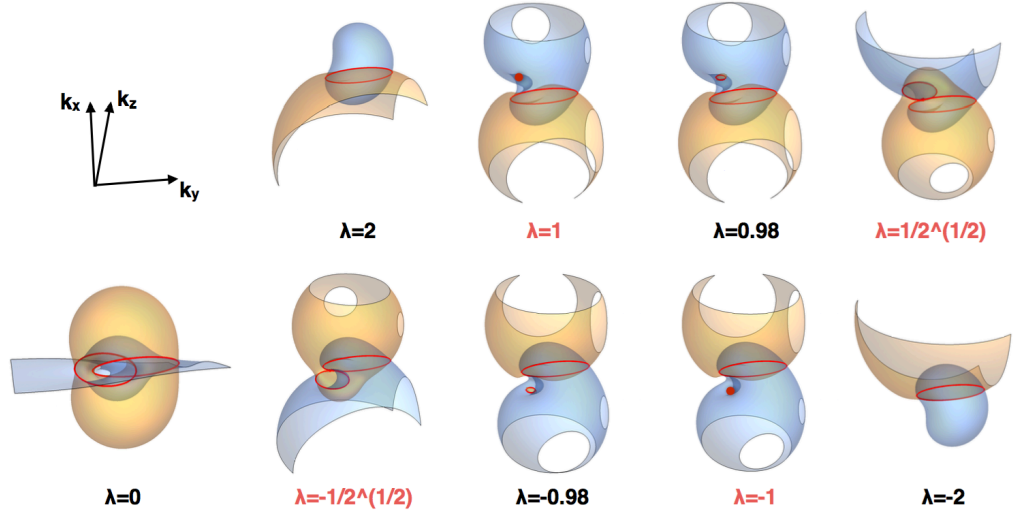

FIG. 13: The nodal line evolution of the Hopf-link model. Here $\lambda$ has four critical points in red color. Notice only $\lambda_c = \pm 1/\sqrt{2}$ are the knot transition points. The $h_x(k) = 0$ and $h_z(k) = 0$ are represented by the blue and yellow surfaces respectively.

In order to determine the phase transition point, we use the $\boldsymbol{T}(\boldsymbol{k}_0) = 0$, that is

$$
\varepsilon_{\alpha\beta\gamma}\hat{\boldsymbol{e}}_\alpha \partial_\beta h_x(\boldsymbol{k}_0)\partial_\gamma h_z(\boldsymbol{k}_0) = 0,
\tag{18}
$$

where $\alpha, \beta, \gamma = x, y, z$. Combined to the condition $h_x(\boldsymbol{k}_0) = h_z(\boldsymbol{k}_0) = 0$, one can obtain the flowing four critical values of the parameter, $\lambda_c = \pm 1/\sqrt{2}, \pm 1$. Only the $\lambda_c = \pm 1/\sqrt{2}$ are the linking transition points and the corresponding TPs are $k_{TP} = (\pm 1/\sqrt{2}, 0, \pm 1/\sqrt{2})$.

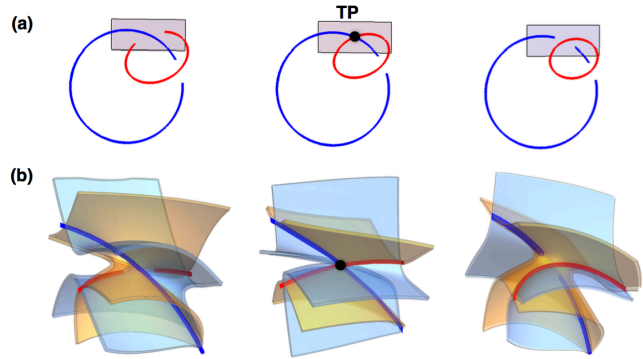

FIG. 14: The nodal line evolution of Hopf-link model defined by Eq. (17) around $\lambda_c = 1/\sqrt{2}$. (a). Nodal lines (red/blue) and natural projective planes (gray) with $\delta\lambda = -0.1, 0, 0.1$ respectively. On the projective plane, the local evolution of the nodal lines belongs to the "crossing-evolution". (b). Local shape of the two surfaces $h_x(k) = 0$ and $h_z(k) = 0$ around the touching point. Both surfaces are in quadratic form near the touching point.

We also plot the local evolutions around the touching points with the following parameters $\lambda_c = 1/\sqrt{2}-0.1, 1/\sqrt{2}, 1/\sqrt{2}+0.1$ in Fig. 14 from the left to right, respectively. It is clear that the Hopf-link transit into two-loop nodal lines as $\delta\lambda$ goes from negative to positive. The local evolution of the nodal lines around the touching point is shown in the natural projective plane in Fig. 14 (a). Around the touching point, $k_{TP} = (1/\sqrt{2}, 0, 1/\sqrt{2})$, One can easily obtain

$$
\begin{aligned}
h_x(\delta k) &\simeq -2\sqrt{2}(\delta k_x \delta k_y + \delta k_y \delta k_z) + 8\delta k_x \delta k_z, \\
h_z(\delta k) &\simeq 3\sqrt{2}(\delta k_x^2 - \delta k_z^2) - 2(\delta k_x \delta k_y - \delta k_y \delta k_z).
\end{aligned}
\tag{19}
$$

Fig. 14 (b) shows the fact that the $h_x(k) = 0$, $h_z(k) = 0$ are both quadratic at the TP and the corresponding local nodal line evolutions.

### B. A four-band model

We also provide a four-band model to investigate the knot transition. The Hamiltonian is

$$
\mathcal{H}(\boldsymbol{k}) = \gamma_{30}(\boldsymbol{k})\Gamma_{30} + \gamma_{32}(\boldsymbol{k})\Gamma_{32} + \gamma_{20}(\boldsymbol{k})\Gamma_{20} + \gamma_{22}(\boldsymbol{k})\Gamma_{22},
\tag{20}
$$

where $\Gamma_{\mu\nu} = \sigma_\mu \otimes \tau_\nu$, $\gamma_{30}(\boldsymbol{k}) = -21/8 + \cos k_x + \cos k_y + \cos k_z$, $\gamma_{32}(\boldsymbol{k}) = 1/2 \sin k_y$, $\gamma_{20}(\boldsymbol{k}) = \lambda + 1/2 \sin k_x$, and $\gamma_{22}(\boldsymbol{k}) = \sin k_z$. The nodal line is determined by

$$
\det[h(\boldsymbol{k})] := \gamma^r(\boldsymbol{k}) + i\gamma^i(\boldsymbol{k}) = [\gamma_{20}(\boldsymbol{k})^2 + \gamma_{30}(\boldsymbol{k})^2 - \gamma_{22}(\boldsymbol{k})^2 - \gamma_{32}(\boldsymbol{k})^2] + 2i[\gamma_{30}(\boldsymbol{k})\gamma_{22}(\boldsymbol{k}) - \gamma_{20}(\boldsymbol{k})\gamma_{32}(\boldsymbol{k})] = 0.
\tag{21}
$$

We show the corresponding nodal line evolution in Fig. 15.

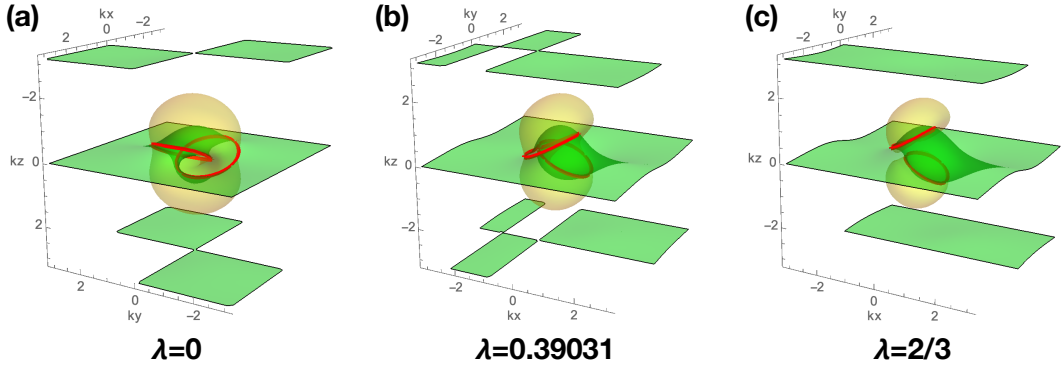

FIG. 15: The nodal line evolution in the four-band model. The phase transition point can be calculated from Eq. 22

The singularity point can be calculated by

$$
\gamma^r(\boldsymbol{k}) = 0, \gamma^i(\boldsymbol{k}) = 0, \vec{\nabla}_{\boldsymbol{k}}\gamma_0^r(\boldsymbol{k}_0) \times \vec{\nabla}_{\boldsymbol{k}}\gamma_0^i(\boldsymbol{k}_0) = 0.
\tag{22}
$$

There exist two phase transition points $\lambda = \pm 0.39031$. Here we only focus on the one shown in Fig. 15. The TP is located at $\boldsymbol{k}_{TP} = (-0.89567, 0, 0)$. Expanding around this point, one can obtain,

$$
\det[h(\boldsymbol{k}_{TP} + \delta\boldsymbol{k})] \simeq (0.71\delta k_x^2 - 0.25\delta k_y^2 - \delta k_z^2) - i(0.31\delta k_x \delta k_y + 1.56\delta k_x \delta k_z),
\tag{23}
$$

which only has quadratic terms.

## VII.  PHYSICAL CONSTRAINT OF LOCAL EVOLUTIONS

In the main text, we have mentioned that the TP evolution can be further classified to the following three cases: (i) the two gradients are parallel ($\vec{\nabla}_{\boldsymbol{k}}\gamma_0^r(\boldsymbol{k}_{\text{TP}}) = c\vec{\nabla}_{\boldsymbol{k}}\gamma_0^i(\boldsymbol{k}_{\text{TP}}) \neq 0$), (ii) one of the gradients vanishes, and (iii) both vanish as shown in Fig. 3 in the main text. Now we will discuss the above constraint to the local evolutions. In order to simplify the discussion, we assume the quadratic terms are nonzero.

We first prove that in the case (i) and (ii), the TP must belong to the Type II. Since in both cases, there always exists a natural projective plane, whose normal vector is given by $\vec{\nabla}_{\boldsymbol{k}}\gamma_0^r(\boldsymbol{k}_{\mathrm{TP}})$ (or $\vec{\nabla}_{\boldsymbol{k}}\gamma_0^i(\boldsymbol{k}_{\mathrm{TP}})$), as shown in Fig. 16 (a1) and (a2). Now if we consider an integral path (chosen to be a circle) crossing the natural projective plane but do not touch the two nodal lines, the winding number must be zero. To explain the vanishing winding number at the TP, we consider case (i) and (ii) separately. First, for case (i), $\vec{\nabla}_{\boldsymbol{k}}\gamma_0^r(\boldsymbol{k}_{\mathrm{TP}})$ parallels $\vec{\nabla}_{\boldsymbol{k}}\gamma_0^i(\boldsymbol{k}_{\mathrm{TP}})$), the integrand in Eq. 6 for winding number does not have any phase winding; hence, the winding number is zero. Second, for case (ii), the integrand in Eq. 6 possesses different orders in the real and imaginary parts (linear and quadratic in $\boldsymbol{k}$) separately; this incompatibility lead to the vanishing winding number. As shown in Fig. 16 (a1) and (a2), the zero winding numbers in the two integral paths imply that the TP must belong to the type II.

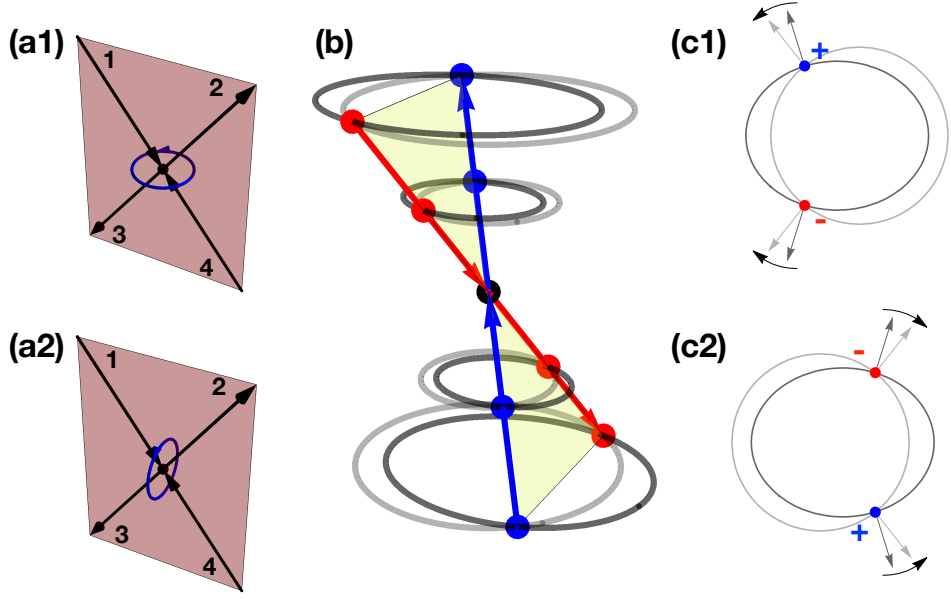

FIG. 16: Physical constraint to the local evolutions are related to the winding number of the nodal lines and TP.

For the second example, starting from $\nabla \gamma_0^r(\boldsymbol{k}_{TP}) = 0$, we can obtain the following expansion

$$\gamma_0^r(\boldsymbol{k}_{TP} + \delta\boldsymbol{k}) = \sum_{i,j} v_{ij}\delta k_i \delta k_j + o(\delta\boldsymbol{k}^3). \tag{24}$$

After taking a proper rotation, $\gamma_0^r(\boldsymbol{k}_{TP} + \delta\boldsymbol{k})$ can be expressed as

$$\gamma_0^r(\boldsymbol{k}_{TP} + \delta\boldsymbol{k}) = v_x \delta\tilde{k}_x^2 + v_y \delta\tilde{k}_y^2 + v_z \delta\tilde{k}_z^2 + o(\delta\boldsymbol{k}^3). \tag{25}$$

The geometry of the equation above can be classified by the following two cases (i) a point ($\mathrm{sign}(v_x) = \mathrm{sign}(v_y) = \mathrm{sign}(v_z)$) and (ii) a "Dirac cone" ($\mathrm{sign}(v_i) = \mathrm{sign}(v_j) = -\mathrm{sign}(v_k)$) . Since only the Dirac cone can form two nodal lines, we focus on the evolution of the Dirac cone. Without loss of generality, we can assume $v_{x/y} > 0$ and $v_z < 0$ in the following discussion. As a result, the possible geometry realization of two nodal lines with a TP is shown in Fig. 3 in the main text. Next we show the TP must belong to the type I. We make two plane cuts above and below the TP as illustrated in Fig. 16 (b) and (c). In Fig. 16 (c), we plot the vectors aligned with $\vec{\nabla}_{\boldsymbol{k}}\gamma_0^{r/i}(\boldsymbol{k})$ at the interaction point (light gray and gray) of the nodal lines. The directions of $\vec{\nabla}_{\boldsymbol{k}}\gamma_0^{r/i}(\boldsymbol{k})$ show that the two intersection points in each plane have the opposite winding numbers by using Eq. 7. One can notice, the orientation of the nodal lines around the TP must be the arrangement shown in Fig. 16 (b). Thus, the TP belongs to the type I TP.

We can take a simple example to illustrate the above results, where the Hamiltonian satisfies $\mathcal{H}_0(\boldsymbol{k}) = h_0(\boldsymbol{k})\tau_+ + h_0^\dagger(\boldsymbol{k})\tau_-$ with $h_0(\boldsymbol{k}) = (k_x^2 + k_y^2 - k_z^2) - i(k_x^2 + 3k_y^2/2 + k_x k_z/2 - k_z^2)$. The two surfaces $\gamma_0^r(\boldsymbol{k}) = 0$ and $\gamma_0^i(\boldsymbol{k}) = 0$ form two "Dirac cones" respectively in the 3D BZ as shown in Fig. 17-i with blue and yellow colors respectively, and the crossing of the two nodal lines forms a TP. We introduce the first type of perturbation in the form of $\mathcal{H}_1(\lambda, \boldsymbol{k}) = \lambda h_1(\lambda, \boldsymbol{k})\tau_+ + \lambda h_1^\dagger(\lambda, \boldsymbol{k})\tau_-$, with $\lambda h_1(\boldsymbol{k}, \lambda) = \lambda[-(\lambda + k_y) + i(\lambda - 2k_y)]$ and $\lambda = 0.1$ for Fig. 17-ii and $\lambda = -0.1$ for -iii. In this regard, when the two "Dirac cones" evolve two cylinders with varying $\lambda$, the two nodal lines on the top cone connect the two on the bottom cone respectively as a crossing evolution. The other type of perturbation is given by $\lambda h_1(\boldsymbol{k}, \lambda) = \lambda$. Fig. 17-iv shows the evolution of these two surfaces and their crossing nodal lines for $\lambda = 0.1$.

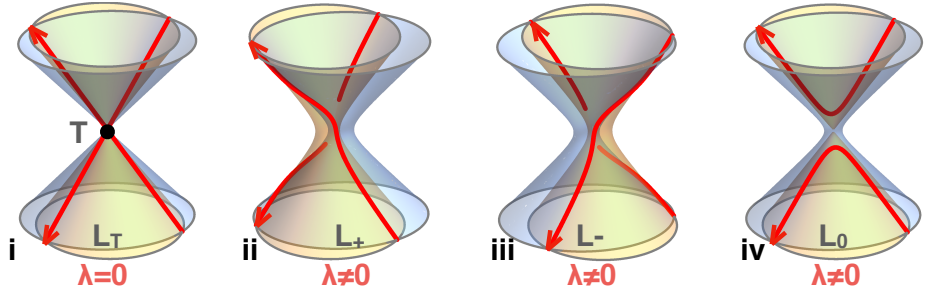

FIG. 17: The local evolution of the nodal lines under different perturbations. Here we choose $h_0(\boldsymbol{k}) = k_x^2 + k_y^2 - k_z^2 - i(k_x^2 + 3k_y^2/2 + k_x k_z/2 - k_z^2)$ for i. The perturbations we chosen for ii/iii is $\lambda h_1(\boldsymbol{k}) = \lambda[-(\lambda + k_y) + i(\lambda - 2k_y)]$ with $\lambda = 0.1$ and $-0.1$ respectively. For iv is $\lambda h_1(\boldsymbol{k}) = \lambda$ with $\lambda = -0.1$.

## VIII. NODAL CHAINS WITH FOUR TPS

In the main text, we have use the following model to study the knot transition,

$$\mathcal{H}_0(\boldsymbol{k}) = h_0(\boldsymbol{k})\tau_+ + h_0^\dagger(\boldsymbol{k})\tau_-, \tag{26}$$

where

$$h_0(\boldsymbol{k}) = 2\cos 2k_x + \cos k_x + 3\cos k_y - 3\cos k_z - 1/10 - 2i\sin k_y \sin k_z \tag{27}$$

Obviously, the nodal chain is determined by

$$\sin k_y \sin k_z = 0, \; 2\cos 2k_x + \cos k_x + 3\cos k_y - 3\cos k_z - 1/10 = 0. \tag{28}$$

This is equivalent to

$$\{k_y = 0, 2\cos 2k_x + \cos k_x - 3\cos k_z + 29/10 = 0\} \cup \{k_y = \pm\pi, 2\cos 2k_x + \cos k_x - 3\cos k_z - 31/10 = 0\}$$
$$\cup \{k_z = 0, 2\cos 2k_x + \cos k_x + 3\cos k_y - 31/10 = 0\} \cup \{k_z = \pm\pi, 2\cos 2k_x + \cos k_x + 3\cos k_y + 29/10 = 0\}. \tag{29}$$

Thus we have two separated nodal chains located at $k_0$ and $k_z = \pm\pi$ planes. In order to simplify the discussion, we only focus on the ones at the $k_z = 0$ plane.

## IX. NODAL CHAINS WITH TWO TPS

In the main text, we have mentioned that the knot transition is forbidden in a nodal chain semimetal protected by two mirror planes with two TPs. Here we show a concrete example to illustrate this. One can consider the following simple example with the Hamiltonian $\mathcal{H}_0(\boldsymbol{k}) = h_0(\boldsymbol{k})\tau_+ + h_0^\dagger(\boldsymbol{k})\tau_-$ and

$$h_0(\boldsymbol{k}) = \cos k_x + \cos k_y - \cos k_z - 1/2 - i\sin k_y \sin k_z. \tag{30}$$

This model has two mirror planes, which are $k_z = 0$ and $k_y = 0$. The nodal chain is shown in Fig. 18 with the red curves. According to our theory, the local evolutions around the two TPs must belong to the types shown in the first row of Fig. 3 in the main text in the viewpoint of natural projective plane (gray and light green squares in Fig. 18). Because we have two touching points, there only exist four possible configurations for the evolutions, as shown in Fig. 18 (a), (b), (c) and (d). All the possible evolutions are either an unknot or an unlink.

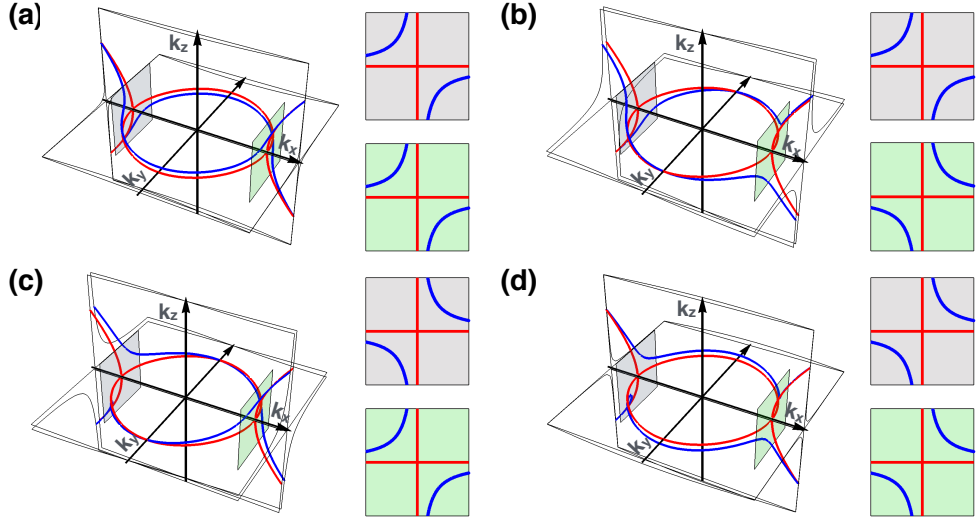

FIG. 18: The red curve shows a nodal chain semimetal protected by two mirror plane symmetries. According to our theory, the local evolutions around the TP can only have two different types as shown on the right, where the two squares are the natural projective planes. Thus global evolution of the nodal chain can only have four classes, as shown in (a), (b), (c) and (d). An unknot or an unlink is the only possible evolutions.

## X. DISCRIMINANT

In the main text, we have mentioned that the degeneracy condition of a general $N$-band non-Hermitian Hamiltonian is determined by the vanishing of the discriminant

$$\Delta_f(\boldsymbol{k}) = \prod_{i<j} \left(E_i(\boldsymbol{k}) - E_j(\boldsymbol{k})\right)^2 = 0, \tag{31}$$

where $\Delta_f(\boldsymbol{k})$ is the discriminant of the characteristic polynomial $f(E, \boldsymbol{k})$ as a function of $E$. Mathematically, the discriminant of a polynomial is defined as follows. Let $f = a_n x^n + ... + a_0$ be a polynomial with coefficients in an arbitrary field $F$. Then the (standard) discriminant of $f$ is defined as

$$\Delta_f := a_n^{2n-2} \Delta_0(f) = a_n^{2n-2} \prod_{1 \le i < j \le n} (\xi_i - \xi_j)^2, \tag{32}$$

where $\xi_1, ..., \xi_n$ are the roots of $f$ in some extension of $F$. For the characteristic polynomial $f(E, \boldsymbol{k})$, the coefficient of $E^N$ is 1, namely $a_N = 1$. This reduces Eq. 32 to Eq. 31. There exist a theorem that relates the discriminant to the Sylvester matrix [4, 5], namely,

$$\Delta_f = (-1)^{n(n-1)/2} a_n^{-1} \det[\mathrm{Syl}(f, f')], \tag{33}$$

where $f' = \partial_E f$ and the Sylvester matrix of two polynomials $f(x) = a_n x^n + ... + a_0$, $g(x) = b_m x^m + ... + b_0 \in F[x]$ is defined by

$$\mathrm{Syl}(f,g) = \begin{pmatrix} a_n & a_{n-1} & a_{n-2} & \dots & 0 & 0 & 0 \\ 0 & a_n & a_{n-1} & \dots & 0 & 0 & 0 \\ \vdots & \vdots & \vdots & & \vdots & \vdots & \vdots \\ 0 & 0 & 0 & \dots & a_1 & a_0 & 0 \\ 0 & 0 & 0 & \dots & a_2 & a_1 & a_0 \\ b_m & b_{m-1} & b_{m-2} & \dots & 0 & 0 & 0 \\ 0 & b_m & b_{m-1} & \dots & 0 & 0 & 0 \\ \vdots & \vdots & \vdots & & \vdots & \vdots & \vdots \\ 0 & 0 & 0 & \dots & b_1 & b_0 & 0 \\ 0 & 0 & 0 & \dots & b_2 & b_1 & b_0 \end{pmatrix}, \tag{34}$$

where $a_n, ..., a_0$ are the coefficients of $f$ and $b_m, ..., b_0$ are the coefficients of $g$. Therefore, if $f$ has a double root in some extension of $F$ if and only if $\Delta_f = 0$. Since the coefficients of the characteristic polynomials are single-valued function of $\boldsymbol{k}$, the corresponding discriminant must also be a single-valued function of $\boldsymbol{k}$.

### A. Some examples

Now we will calculate some examples.

#### 1. $n = 2$ case

If $f(x) = ax^2 + bx + c$, then

$$\Delta_f = -a^{-1}R(f, f') = -a^{-1}\det\begin{pmatrix} a & b & c \\ 2a & b & 0 \\ 0 & 2a & b \end{pmatrix} = b^2 - 4ac. \tag{35}$$

#### 2. $n = 3$ case

If $f(x) = ax^3 + bx^2 + cx + d$, then

$$\Delta_f = -a^{-1}R(f, f') = -a^{-1}\det\begin{pmatrix} a & b & c & d & 0 \\ 0 & a & b & c & d \\ 0 & a & b & c & d \\ 0 & 3a & 2b & c & 0 \\ 0 & 0 & 3a & 2b & c \end{pmatrix} \tag{36}$$
$$= b^2c^2 - 4ac^3 - 4b^3d + 18abcd - 27a^2d^2.$$

### B. Application to non-Hermitian systems

In this section, we will calculate the non-Hermitian degenerate points in two- and three- band models based on the discriminant. We want to emphasize that this method can only be applied to calculate the non-Hermitian degeneracies, since the energy spectrum is extended to the complex field.

#### 1. Two-band example

For a general two-band model,

$$\mathcal{H}(\boldsymbol{k}) = h_0(\boldsymbol{k}) + h_x(\boldsymbol{k})\sigma_x + h_y(\boldsymbol{k})\sigma_y + h_z(\boldsymbol{k})\sigma_z. \tag{37}$$

The characteristic equation of the two band model can be written as

$$f(E, \boldsymbol{k}) = E^2 + b(\boldsymbol{k})E + c(\boldsymbol{k}), \tag{38}$$

where $b(\boldsymbol{k}) = -2h_0(\boldsymbol{k})$ and $c(\boldsymbol{k}) = h_0^2(\boldsymbol{k}) - h_x^2(\boldsymbol{k}) - h_y^2(\boldsymbol{k}) - h_z^2(\boldsymbol{k})$ with $h_\mu(\boldsymbol{k}) = h_\mu^r(\boldsymbol{k}) + ih_\mu^i(\boldsymbol{k})$ as complex function of $\boldsymbol{k}$. The discriminant of Eq. 38 to the variable $E$ is

$$\Delta_f(\boldsymbol{k}) = b^2(\boldsymbol{k}) - 4c(\boldsymbol{k}) = 4[h_x^2(\boldsymbol{k}) + h_y^2(\boldsymbol{k}) + h_z^2(\boldsymbol{k})] = 0. \tag{39}$$

Obviously, this is indeed the band degeneracy condition for the two-band model.

## 2. Three-band model

Since the term proportional to the identity matrix can not affect the band degenracies, the general three-band model with out $I_{3\times3}$ term can be written as,

$$\mathcal{H}(\boldsymbol{k}) = \sum_{\rho=1}^{8} g_\rho(\boldsymbol{k})\lambda_\rho, \tag{40}$$

where the eight Gell-Mann matrices are

$$\lambda_1 = \begin{pmatrix} 0 & 1 & 0 \\ 1 & 0 & 0 \\ 0 & 0 & 0 \end{pmatrix}, \quad \lambda_2 = \begin{pmatrix} 0 & -i & 0 \\ i & 0 & 0 \\ 0 & 0 & 0 \end{pmatrix}, \quad \lambda_3 = \begin{pmatrix} 1 & 0 & 0 \\ 0 & -1 & 0 \\ 0 & 0 & 0 \end{pmatrix}, \quad \lambda_4 = \begin{pmatrix} 0 & 0 & 1 \\ 0 & 0 & 0 \\ 1 & 0 & 0 \end{pmatrix}, \tag{41}$$

$$\lambda_5 = \begin{pmatrix} 0 & 0 & -i \\ 0 & 0 & 0 \\ i & 0 & 0 \end{pmatrix}, \quad \lambda_6 = \begin{pmatrix} 0 & 0 & 0 \\ 0 & 0 & 1 \\ 0 & 1 & 0 \end{pmatrix}, \quad \lambda_7 = \begin{pmatrix} 0 & 0 & 0 \\ 0 & 0 & -i \\ 0 & i & 0 \end{pmatrix}, \quad \lambda_8 = \frac{1}{\sqrt{3}} \begin{pmatrix} 1 & 0 & 0 \\ 0 & 1 & 0 \\ 0 & 0 & -2 \end{pmatrix}. \tag{42}$$

The characteristic equation can be written as

$$f(E, \boldsymbol{k}) = E^3 + c(\boldsymbol{k})E + d(\boldsymbol{k}) = 0, \tag{43}$$

where $c = -\sum_{s=1}^{8} g_s^2$ and $d = g_8 \left(-6g_1^2 - 6g_2^2 - 6g_3^2 + 2g_8^2 + 3\left(g_4^2 + g_5^2 + g_6^2 + g_7^2\right)\right)/3^{3/2} - 2g_1(g_4g_6 + g_5g_7) + 2g_2(g_4g_7 - g_5g_6) + g_3\left(-g_4^2 - g_5^2 + g_6^2 + g_7^2\right)$. The discriminant of Eq. 43 to the variable $E$ is

$$\Delta_f(\boldsymbol{k}) = -4c^3(\boldsymbol{k}) - 27d^2(\boldsymbol{k}) = 0. \tag{44}$$

This is a quite simple equation, which can be analytical dealt in 3D systems.

---



\end{document}